\documentclass[a4paper,twocolumn,pra,aps,nofootinbib,showpacs]{revtex4}
\newcommand{\beq}{\begin{equation}}
\newcommand{\eeq}{\end{equation}}
\newcommand{\beqa}{\begin{eqnarray}}
\newcommand{\eeqa}{\end{eqnarray}}

\def\ra{\rangle}
\def\la{\langle}

\usepackage{amsmath,amsfonts,amssymb}
\usepackage{graphicx}
\DeclareGraphicsExtensions{.pdf,.png,.jpg}
\usepackage{epstopdf}
\usepackage{bm}
\usepackage{dsfont}

\usepackage{color}

\begin{document}
\title{Fast expansions and compressions of trapped-ion chains}
\author{M. Palmero$^{1}$}
\author{S. Mart\'inez-Garaot$^{1}$}
\author{J. Alonso$^{2}$}
\author{J. P. Home $^{2}$}
\author{J. G. Muga$^{1,3}$}
\affiliation{$^{1}$Departamento de Qu\'{\i}mica F\'{\i}sica, Universidad del Pa\'{\i}s Vasco - Euskal Herriko Unibertsitatea,
Apdo. 644, Bilbao, Spain}
\affiliation{$^{2}$Institute for Quantum Electronics, ETH Z\"urich, Otto-Stern-Weg 1, 8093 Z\"urich, Switzerland}
\affiliation{$^{3}$Department of Physics, Shanghai University, 200444 Shanghai, People's Republic of China}
\begin{abstract}
We investigate the dynamics under diabatic expansions/compressions of linear ion chains.
%For diabatic trap frequency drivings,
%we study how each motional degree of freedom is excited depending on the composition of the  ion chain.
Combining a dynamical normal-mode harmonic approximation with the invariant-based inverse-engineering technique, we design protocols that minimize the final motional excitation of the ions. This  can substantially reduce the transition time between high and low trap-frequency operations, potentially contributing to the development of scalable quantum information processing.
\end{abstract}
\maketitle\maketitle
\section{Introduction}
Trapped-ion chains provide currently a leading architecture to test and develop quantum information processing protocols \cite{08Blatt,reviewsim1,reviewsim2}. A toolbox of fundamental ion manipulations is required to implement the logical operations or simulations. This comprises procedures to control the ions' internal (electronic) and external (motional) degrees of freedom, and also to couple them together. Such procedures should be not only feasible but also fast. Achieving high speeds is important in itself, to allow for more operations per unit time, but it is also instrumental in suppressing the effects of noise and decoherence.

Regarding motional control, a number of operations have been identified as relevant for different scalable trapped-ion quantum information architectures \cite{02Kielpinski,00Cirac,14Monroe}, including ion transport \cite{Erik,Palmero,Palmero2,Bowler,Walther,13Alonso}, and ion-chain splitting and recombination \cite{Bowler,11Home,Kaufmann,Ruster}. These operations can be performed on single or mixed-species ion chains \cite{13Home}, allowing for sympathetic ion cooling or quantum-logic spectroscopy \cite{05Schmidt}.

The physical operation that we consider here is fast control of the motional frequencies of the trapped ions, which in the case of multiple ions leads to chain expansions and compressions. Several elementary protocols benefit from a high trap-frequency whereas others are better performed with low frequencies. Therefore, a fast transition between them without inducing final excitations is a worthwhile goal.

Operations that benefit from high motional frequencies (\emph{i.e.} large potential curvature, small inter-ion distance, and small Lamb-Dicke parameters) include:

\begin{itemize}
\item{Doppler laser cooling, since the mean phonon number is lower for tighter traps \cite{86Stenholm};}
\item{any operation where a single motional normal mode (NM) of an ion chain needs to be spectrally resolved, since the NM frequency splitting is proportional to the trap curvature \cite{11Home};}
\item{operations which make use of motional sidebands and whose fidelity is limited by off-resonantly driving carrier transitions on the qubit.}
%\item{high frequencies are generally advantageous also considering that motional heating rates due to noise on the trapping potential are typically proportional to the inverse frequency squared \cite{00Turchette,11Danilidis}.}
\end{itemize}
On the other hand, operations where a lower motional frequency is desired include:
\begin{itemize}
\item{single-ion addressing in a multi-ion crystal;}
\item{resolved sideband cooling, which cools at a rate proportional to the square of the Lamb-Dicke parameter \cite{03Leibfried2};}
\item{geometric phase gates \cite{03Leibfried}, which are faster for larger Lamb-Dicke parameters.}
\end{itemize}
In many cases a compromise will be optimal, depending on the dominant limitations for a particular experiment.

%For one particle, or a Bose-Einstein condensate (BEC), 
Fast expansions/compressions without final excitation have been designed in a number of different ways \cite{Udo,Koslof,Masuda,Muga,Xi,Review,Adolfo}. Invariant-based engineering or scaling methods \cite{Muga,Xi} were realized experimentally for a non-interacting cold-atom cloud \cite{Nice1} and a Bose-Einstein condensate \cite{Nice1,Nice2}.  However, the methods used rely on single-particles, 
BEC dynamics, or equal masses, and
are not directly applicable to an arbitrary interacting ion chain. We propose here a method to design trap expansions and compressions faster than adiabatically and without final motional excitation. Specifically we define dynamical normal modes similar to the ones defined for shuttling ion chains in \cite{Palmero} and apply invariant-based inverse-engineering techniques by either exact or approximate methods.

We first discuss two-ion chains in Sec. \ref{2ion}, both for ions of equal mass, and ions of different mass, and then extend the analysis in Sec. \ref{Nion} to longer chains. In the examples only expansions of the trapping potential are considered, as
compressions may be performed with the same time-evolution of the spring constant, only time-reversed.
\section{2-ion chain expansion\label{2ion}}
We will deal with a one-dimensional trap containing an $N$-ion chain whose Hamiltonian in terms of the
positions $\{q_i\}$ and momenta $\{p_i\}$ of the ions in the laboratory frame is
\beq
\label{exphamiltonian}
H=\sum_{i=1}^N\frac{p_i^2}{2m_i}+\sum_{i=1}^N\frac{1}{2}u_0(t)q_i^2+\sum_{i=1}^{N-1}\sum_{j=i+1}^N\frac{C_c}{q_i-q_j},
\eeq
where $C_c=\frac{e^2}{4\pi\epsilon_0}$, with $\epsilon_0$ the vacuum permittivity. $u_0(t)$ is the common (time-dependent) spring constant that defines the oscillation frequencies $\omega_j(t)/(2\pi)$ for the different ions in the absence of Coulomb coupling: $u_0(t)=m_1\omega^2_1(t)=m_2\omega^2_2(t)=\dots=m_N\omega^2_N(t)$. All ions are assumed to have the same charge $e$, and be ordered as  $q_1>q_2>\dots>q_N$, with negligible overlap of probability densities as a result of the Coulomb repulsion.
The potential term $V(q_1,q_2)=H-\sum_ip_i^2/2m_i$ in the Hamiltonian (\ref{exphamiltonian}) for two ions is minimal at
the equilibrium points $q_1^{(0)}=x_0/2$, $q_2^{(0)}=-x_0/2$, where $x_0\equiv x_0(t)~=~2\scriptsize{\sqrt[3]{\frac{C_c}{4u_0(t)}}}$
is the equilibrium distance between the two ions.
Instantaneous, mass-weighted, NM coordinates are defined by diagonalizing the matrix
$V_{ij}=\frac{1}{\sqrt{m_im_j}}\frac{\partial^2V}{\partial q_i\partial q_j}(q_i^{(0)},q_j^{(0)})$ \cite{Palmero}.
The time-dependent eigenvalues are \cite{01Morigi}
\beq
\lambda_\pm=\left(1+\frac{1}{\mu}\pm\sqrt{1-\frac{1}{\mu}+\frac{1}{\mu^2}}\right)\omega_1^2,
\eeq
where we have relabeled $m_1\rightarrow m$ and $m_2\rightarrow \mu m$, and omitted the explicit time dependences
to avoid a cumbersome notation, \emph{i.e.}, 
$\lambda_\pm\equiv\lambda_\pm(t)$ and $\omega_1\equiv\omega_1(t)$.  
The time-dependent angular frequencies for each mode are
\beq
\label{ome}
\Omega_\pm\equiv\Omega_\pm(t)=\sqrt{\lambda_\pm}=
\left(1+\frac{1}{\mu}\pm\sqrt{1-\frac{1}{\mu}+\frac{1}{\mu^2}}\right)^{1/2}\!\! \omega_1,
\eeq
and the eigenvectors corresponding to these eigenvalues are $v_\pm=(a_\pm,b_\pm)^T$,
where
\beqa
\label{coefficients}
a_+&=&\left(\frac{1}{1+\left(1-\frac{1}{\mu}-\sqrt{1-\frac{1}{\mu}+\frac{1}{\mu ^2}}\right)^2\mu}\right)^{1/2},
\nonumber\\
b_+&=&\left(1-\frac{1}{\mu}-\sqrt{1-\frac{1}{\mu}+\frac{1}{\mu ^2}}\right)\sqrt{\mu}a_+,
\nonumber\\
a_-&=&\left(\frac{1}{1+\left(1-\frac{1}{\mu}+\sqrt{1-\frac{1}{\mu}+\frac{1}{\mu ^2}}\right)^2\mu}\right)^{1/2},
\nonumber\\
b_-&=&\left(1-\frac{1}{\mu}+\sqrt{1-\frac{1}{\mu}+\frac{1}{\mu ^2}}\right)\sqrt{\mu}a_-.
\eeqa
The instantaneous, dynamical normal-mode (mass weighted) coordinates are finally
\beqa
{\sf q}_+&=&a_+\sqrt{m}\left(q_1-\frac{x_0}{2}\right)+b_+\sqrt{\mu m}\left(q_2+\frac{x_0}{2}\right),
\nonumber\\
{\sf q}_-&=&a_-\sqrt{m}\left(q_1-\frac{x_0}{2}\right)+b_-\sqrt{\mu m}\left(q_2+\frac{x_0}{2}\right).
\eeqa
The quantum dynamics of a state $|\psi\ra$ governed by $H$ in the laboratory frame
may be transformed into  the  moving frame of NM coordinates
by the unitary operator
\beq
\label{unitary}
U= \int d{\sf q}_+d{\sf q}_-dq_1dq_2 |{\sf q}_+,{\sf q}_-\rangle\langle {\sf q}_+,{\sf q}_-| q_1,q_2\rangle \langle q_1,q_2|,
\eeq
where  $\langle {\sf q}_+,{\sf q}_-|q_1,q_2\rangle =\delta [q_1-q_1({\sf q}_+,{\sf q}_-)]\delta [q_2-q_2({\sf q}_+,{\sf q}_-)]$.
%\equiv\delta_1\delta_2$.
The Hamiltonian in the dynamical equation for $|\psi'\ra=U|\psi\ra$ is given by
\beqa
\label{Hprime}
H'&=&UHU^\dagger -i\hbar U(\partial_t U^\dagger)=\nonumber\\
&=&\sum_\nu\left(\frac{{\sf p}_\nu^2}{2}-{\sf p}_{0\nu}{\sf p}_\nu+\frac{1}{2}\Omega_\nu^2{\sf q}_\nu^2\right),
\eeqa
where cubic and higher order terms in the coordinates have been neglected, $\nu=\pm$,  ${\sf p}_\pm$ are (mass weighted) momenta conjugate to ${\sf q}_\pm$, and
\beq
\label{p0}
{\sf p}_{0\pm}=-\dot{{\sf q}}_\pm=\frac{2}{3}(-a_\pm \sqrt{m_1}+b_\pm\sqrt{m_2})\sqrt[3]
{\frac{C_c}{4m_1\omega_1^5}}\dot{\omega}_1
\eeq
are functions of time with the same dimensions as the mass weighted momenta. They appear because of the time dependence of the NM coordinates through $x_0$, which is a function of $\omega_1(t)$.
These ${\sf p}_{0\pm}$ functions act as momentum shifts
in a further unitary transformation which suppresses the terms linear in ${\sf p}_\pm$,
\beqa
\label{hamNM}
\mathcal{U}&=&e^{-i({\sf p}_{0+}{\sf q}_++{\sf p}_{0-}{\sf q}_-)/\hbar},\nonumber\\
|\psi''\ra&=&\mathcal{U}|\psi'\ra,\nonumber\\
H''&=&\mathcal{U}H'\mathcal{U}^\dagger -i\hbar \mathcal{U}(\partial_t\mathcal{U}^\dagger)=\nonumber\\
   &=&\sum_\nu \left[\frac{{\sf p}_\nu^2}{2}+\frac{1}{2}\Omega_\nu\left({\sf q}_\nu +\frac{\dot{{\sf p}}_{0\nu}}{\Omega_\nu^2}\right)^2\right].
\eeqa
This Hamiltonian corresponds to two effective harmonic oscillators with time-dependent frequencies and a time-dependent moving center. Note that the ``motion'' of the harmonic oscillators is in the normal-mode-coordinate  space,
and that the actual center of the external trap in the laboratory frame is fixed. According to Eqs. (\ref{ome}) and (\ref{p0}) both the NM harmonic oscillators' centers  ($-{\dot{{\sf p}}_{0\pm}}/{\Omega_\pm^2}$)
and the frequencies ($\Omega_\pm$) depend on $\omega_1(t)$. This is important as, to solve the dynamics for given $\omega_1(t)$, the oscillators are effectively independent. However, from an inverse-engineering perspective,  their time-dependent parameters cannot be designed independently. This ``coupling'' is here more involved than for the transport of two ions in a rigidly moving harmonic trap \cite{Palmero}, where ${\sf p}_{0\pm}(t)$ take different forms which depend on the trap position but not on the trap frequency. A different approach is thus required.

The Lewis-Riesenfeld invariants \cite{LR} of the two oscillators are
\beqa
I_\pm&=&\frac{1}{2}[\rho_\pm ({\sf p}_\pm-\dot{\alpha}_\pm)-\dot{\rho}_\pm({\sf q}_\pm-\alpha_\pm)]^2
\nonumber\\
&+&\frac{1}{2}\Omega_{0\pm}^2\left(\frac{{\sf q}_\pm-\alpha_\pm}{\rho_\pm}\right)^2,
\eeqa
where $\Omega_{0\pm}=\Omega_\pm(0)$.
The invariants depend on the  auxiliary functions $\rho_\pm$ (scaling factors of the expansion modes)  and $\alpha_\pm$
(mass scaled centers of the dynamical modes of the invariant).
They satisfy the  auxiliary  (Ermakov and Newton) equations
\beqa
\label{auxiliaryexp1}
\ddot{\rho}_\pm +\Omega_\pm^2\rho_\pm =\frac{\Omega_{0\pm}^2}{\rho_\pm^3},
\\
\label{auxiliaryexp2}
\ddot{\alpha}_\pm +\Omega_{\pm}^2\alpha_\pm =\dot{{\sf p}}_{0\pm}.
\eeqa
Dynamical expansion modes $|\psi_{n\pm}''\ra$ (not to be confused with normal modes) may be found. These are exact time-dependent
solutions of the Schr\"odinger equation and also instantaneous eigenstates of the invariant \cite{Erik},
\beq
\la {\sf q}_\pm|\psi'' _{n\pm}\rangle=e^{\frac{i}{\hbar} \left[\frac{\dot{\rho}_\pm{\sf q}_\pm^2}{2\rho_\pm}+(\dot{\alpha}_\pm\rho_\pm-\alpha_\pm\dot{\rho}_\pm)
\frac{{\sf q}_\pm}{\rho_\pm}\right]}
\frac{\Phi_n(\sigma_\pm)}{\rho_\pm^{1/2}},
\eeq
where $\sigma_\pm=\frac{{\sf q}_\pm-\alpha_\pm}{\rho_\pm}$ and $\Phi_n(\sigma_\pm)$ are the eigenfunctions of the static harmonic
oscillator at time $t=0$. Within  the harmonic approximation the NM wave functions $|\psi''_\pm\ra$ evolve independently with $H''$.
They may be written as combinations of the expansion modes, $|\psi''_{\pm}(t)\ra=\sum_n c_{n\pm} |\psi''_{n\pm}\ra$ with normalized constant amplitudes.
The average energies of  the $n$-th expansion mode for two NM are
\beqa
\label{energy}
E''_{n\pm}&=&\langle \psi''_{n\pm} |H''|\psi''_{n\pm} \rangle
\nonumber\\
&=&\frac{(2n+1)\hbar}{4\Omega_{0\pm}}\left(\dot{\rho}_\pm^2+\Omega_\pm^2\rho_\pm^2+\frac{\Omega_{0\pm}^2}{\rho_\pm^2}\right)\nonumber\\
&+&\frac{1}{2}\dot{\alpha}_\pm^2+\frac{1}{2}\Omega_\pm^2(\alpha_\pm-\dot{{\sf p}}_{0\pm}/\Omega_{0\pm}^2)^2.
\eeqa
In numerical examples the initial ground state is, in the harmonic approximation,  of the form $|\psi''_{0+}(0)\ra|\psi''_{0-}(0)\ra$, so the time dependent
energy is given by $E''(t)=E''_{0+}+E''_{0-}$.
Note that if we impose both unitary operators $U(t)$ and ${\cal U}(t)$ to be 1 at $t=0$ and $t_f$,
the transformed wave function $|\psi''\rangle$ and the laboratory wave function $|\psi\rangle$ will be the same at both these times and the energy
$E''(t=0,t_f)$ will be the same as the laboratory-frame energy. Both unitary transformations satisfy this provided that
$\dot{\omega}_1(t_b)=0$, where $t_b=0, t_f$, as long as the quadratic approximation in the Hamiltonian (\ref{hamNM}) is valid.
%In all numerical examples we shall deal with $n=0$ and with initial boundary conditions that
%make $|\psi_0''(0)\ra=|\psi''_{0+}(0)\ra|\otimes \psi''_{0-}(0)\ra$ the ground state of the initial Hamiltonian.

%
%
%
%
% % % % % % % % % % % % % % % % % % % % % % % % % % % % % % % % % % % % % % % % %
% % % % % % % % % % % % % % % % % % % % % % % % % % % % % % % % % % % % % % % %
\begin{figure}[t]
\begin{center}
\includegraphics[width=8cm]{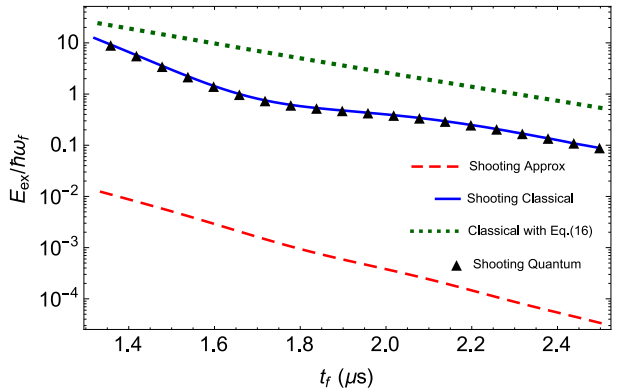}
\caption{\label{optexc}(Color online)
Final excitation energy at $t_f$ for the expansion of two $^{40}$Ca$^+$ ions, with respect to the final ground state (quantum) 
or the final equilibrium energy (classical). The initial state is the ground state (quantum) or the equilibrium state (classical) for the initial trap. 
The dashed red line is the excitation in the harmonic approximation, using Eq. (\ref{energy}) for the NM energies, 
with the protocol obtained by the shooting method; the solid blue line (classical) and black triangles (quantum) are for the same protocol but with the dynamics driven by the full Hamiltonian (\ref{exphamiltonian}).
The dotted green line is for the protocol (\ref{omega}) with the full Hamiltonian.
The parameters used are $\omega_0 /(2\pi)=1.2$ MHz and $\gamma^2=3$.}
\end{center}
\end{figure}
% % % % % % % % % % % % % % % % % % % % % % % % % % % % % % % % % % % % % % % % % % %
% % % % % % % % % % % % % % % % % % % % % % % % % % % % % % % % % % % % % % % % % % %
%

For a single  harmonic oscillator without the independent term in Eq. (\ref{auxiliaryexp2}),
\emph{i.e.} with  a fixed center,
the frequency in a trap expansion
was already inverse engineered in \cite{Xi}.
For this case we can use the same notation as before but no subindices for the auxiliary functions.
$\alpha$ is zero for all times, and in the Ermakov equation
the conditions $\rho(0)=1$, $\rho(t_f)=\gamma=\sqrt{\omega_0/\omega_f}$,
and $\dot\rho(t_b)=\ddot\rho(t_b)=0$,
suffice to avoid any excitation (since $[H(t_b),I(t_b)]=0$) and ensure continuity of the oscillator frequency.
Any interpolated function $\rho(t)$ satisfying these conditions provides a valid $\Omega(t)$.
Similarly, in harmonic transport of an ion (with the trap moving rigidly from 0 to $d$ with  a constant frequency \cite{Erik})
the  auxiliary equation for $\rho$ becomes
trivially satisfied by $\rho=1$ and, to avoid excitations and ensure continuity, $\alpha$ may be any interpolated function satisfying
$\alpha(0)=0$, $\alpha(t_f)=d$, $\dot{\alpha}(t_b)=\ddot{\alpha}(t_b)=0$ \cite{Erik}.
Instead of these simpler settings, when inverse engineering the expansion of the ion chain the auxiliary equations (\ref{auxiliaryexp1}) are
non-trivially coupled
and have to be solved
consistently with Eq. (\ref{auxiliaryexp2}), since
$\Omega_\pm$ and ${\sf p}_{0\pm}$ are functions of the same frequency $\omega_1$. In other words, only interpolated
auxiliary functions $\rho_\pm(t)$, $\alpha_\pm(t)$ consistent with the same $\omega_1(t)$ are valid.
%The problem is further complicated by the second mode. Again, the same $\omega_1$ should be extracted
%from the inversion.

For both NM,
%we obtain boundary conditions using the definition of $\Omega_+$, $\Omega_-$. For both modes
we impose for Eq. (\ref{auxiliaryexp1}) the boundary conditions (BC) $\rho_\pm(0)=1$,
$\rho_\pm(t_f)=\gamma$, $\dot{\rho}_\pm (t_b)=\ddot{\rho}_\pm (t_b)=0$.
Here $\omega_0=\omega_1(0)$ and $\omega_f=\omega_1(t_f)$. 
The BC for the second set of equations are
$\alpha_\pm (t_b)=\dot{\alpha}_\pm(t_b)=\ddot{\alpha}_\pm (t_b)=0$.
Eq. (\ref{auxiliaryexp2}) with Eq. (\ref{p0}) implies that at the boundaries we must have
$\frac{5}{3}\frac{\dot{\omega}_1^2(t_b)}{w_1(t_b)}-\ddot{\omega}_1(t_b)=0$.
This is satisfied by imposing
$\dot{\omega}_1(t_b)=0$, $\ddot{\omega}_1(t_b)=0$.
Substituting these conditions in Eq. (\ref{auxiliaryexp1}) we finally get the extra BC
$\rho_\pm^{(3)}(t_b)=\rho_\pm^{(4)}(t_b)=0$.

% along with those mentioned above.
%Similar to what we did in the transport problem \cite{Palmero},
%we tried to interpolate an ansatz function of $\rho$ that satisfies both auxiliary equations at the same time.
%The idea is to try an ansatz for, for example $\rho_+$ leaving a number of free parameters. Then,
%by solving its corresponding auxiliary equation in (\ref{auxiliaryexp1}), we would get an expression for
%$\omega_1(t)$ depending on this two parameters. Now, substituting this function in the other equation
%in (\ref{auxiliaryexp1}) we would inversely solve for $\rho_-$. Forcing the solution to satisfy all the required
%BC, we would also fix the parameters we left free in the original ansatz.
%However, after solving the first auxiliary equation for the $`+'$ mode, it is beyond our computational
%power to inversely solve the equation for the other mode.

%
%
% % % % % % % % % % % % % % % % % % % % % % % % % % % % % % % % % % % % % % % % %
% % % % % % % % % % % % % % % % % % % % % % % % % % % % % % % % % % % % % % % %
\begin{figure}[t]
\begin{center}
\includegraphics[width=8cm]{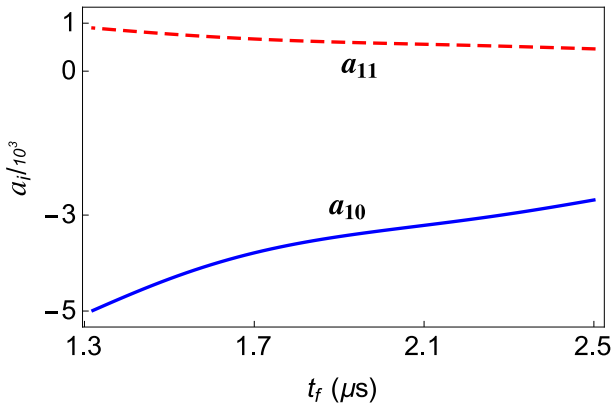}
\caption{\label{freeparameters}(Color online)
The values of the optimizing free parameters in the two-ion expansion $a_{10}$ (solid blue line) and $a_{11}$ (dashed red line) in the expansion of two $^{40}$Ca$^+$ ions starting in the ground state.
$\omega_0 /(2\pi)=1.2$ MHz, $\gamma^2=3$.}
\end{center}
\end{figure}
% % % % % % % % % % % % % % % % % % % % % % % % % % % % % % % % % % % % % % % % % % %
% % % % % % % % % % % % % % % % % % % % % % % % % % % % % % % % % % % % % % % % % % %
%

To engineer the auxiliary functions we proceed as follows:
first we design $\rho_-(t)$\footnote{We choose  $\rho_-$ instead of $\rho_+$ since
$\Omega_+ >\Omega_-$. The effective trap for the plus ($+$) mode is thus tighter and
less prone to excitation than the minus ($-$) mode. Designing first $\rho_-$ guarantees that this `weakest' minus mode will not be excited.}
so as to satisfy the 10  BC for $\rho_-(t_b)$ and their derivatives.
They could be satisfied with a ninth-order polynomial,
but we shall use  higher order polynomials so that free parameters are left.  These
may be chosen to satisfy the equations for the
remaining BC for $\alpha_\pm$ and $\rho_+$.
$\omega_1(t)$ is deduced from the polynomial using Eq. (\ref{auxiliaryexp1}) so it becomes
a function of the free parameters.
There are different ways to fix the free parameters so as to satisfy the remaining BC and design the other auxiliary functions.
In practice we have used a shooting method \cite{shooting}.
The BC used for the shooting are $\alpha_\pm(0)=\dot{\alpha}_\pm(0)=\dot{\rho}_+=0$ and $\rho_+(0)=1$.
Note that if $\alpha_\pm(t_b)=0$, then $\ddot{\alpha}_\pm(t_b)=0$
since  we impose $\dot{\omega}_1(t_b)=\ddot{\omega}_1(t_b)=0$.
The differential equations (\ref{auxiliaryexp1}) for $\rho_+(t)$   and (\ref{auxiliaryexp2}) for $\alpha_\pm$ are now
solved forward in time.

In the following one must distinguish between single-species and mixed-species ion chains. A consequence of having equal mass ions is that $\alpha_-(t)$ is 0 at all times (because the ion chain is symmetric, and thus the center of mass remains static) so we only have to design the three auxiliary functions $\rho_\pm(t)$ and $\alpha_+(t)$. When both ions are of different species, the chain is not symmetric anymore, so we also need to design $\alpha_-$ taking into account its BC.

The MatLab function `fminsearch' \cite{shooting} is used to find the free parameters that minimize the
total final energy for the approximate Hamiltonian, $E''_{0+}(t_f)+E''_{0-}(t_f)$, see Eq. (\ref{energy}).
For equal mass ions, an 11-th order polynomial $\rho_-(t)=\sum_{n=0}^{11}a_nt^n/t_f^n$, \emph{i.e.} two free parameters, is enough
to achieve negligible excitation in a range of times for which the harmonic approximation is valid. Only two free parameters are needed to satisfy the BC $\alpha_+(t_f)=\dot{\alpha}_+(t_f)=0$, whereas $\rho_+(t_f)=\gamma$ is also nearly satisfied for all values of these free parameters
because the evolution of this scaling factor is close to being adiabatic.
$\omega_1(t)$  is then a function of the free parameters $a_{10},a_{11}$. Fig. \ref{optexc}  depicts the final excitation energy for optimized parameters
in the harmonic approximation, using Eq. (\ref{hamNM}), and with the full Hamiltonian (\ref{exphamiltonian}), whereas in Fig. \ref{freeparameters} the values of the optimizing free parameters are represented. The quantum simulations (triangles in Fig. \ref{optexc}) are performed starting from the
ground state of the Hamiltonian (\ref{exphamiltonian}) at $t=0$, which is calculated numerically. 
For the corresponding classical simulations we solve Hamilton's equations for the two ions in the laboratory frame with Eq. (\ref{exphamiltonian}):
the excitation energy is calculated 
as the total energy minus the minimal energy of the ions in equilibrium. The initial conditions correspond 
as well to the ions in equilibrium.  
As the potential is effectively nearly harmonic and the evolution of wave packet's width ($\rho_\pm$) is close to being adiabatic,
the  classical excitation energy reproduces accurately the quantum excitation energy, as demonstrated in Fig. \ref{optexc}.  
Quantum calculations are very demanding, in particular with three or more ions, so that we shall only perform classical calculations from now on. 
%For the final times involved in the calculations, we do not expect squeezing to play a role any larger than in Fig. \ref{optexc}, so we expect the classical calculations to reproduce accurately the quantum ones.

%
%
%
%
% % % % % % % % % % % % % % % % % % % % % % % % % % % % % % % % % % % % % % % % %
% % % % % % % % % % % % % % % % % % % % % % % % % % % % % % % % % % % % % % % %
\begin{figure}[t]
\begin{center}
\includegraphics[width=8.4cm]{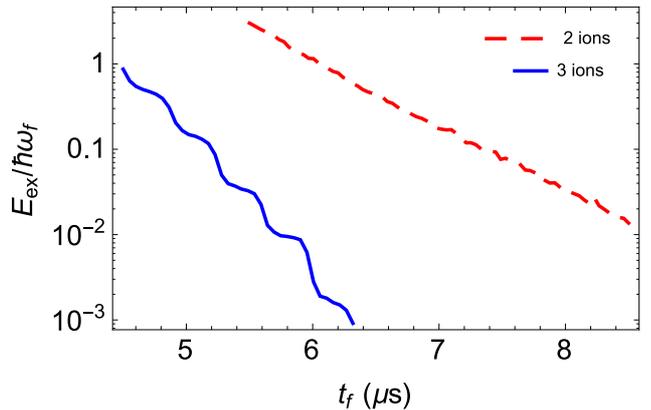}
\caption{\label{excdiff}(Color online)
Final excitation energy for the expansion of a $^{9}$Be$^+$-$^{40}$Ca$^+$ ion chain (solid blue line) and a $^9$Be$^+$-$^{40}$Ca$^+$-$^9$Be$^+$ chain (dashed red line) starting in the equilibrium configuration. The protocols are optimized with four (for $^{9}$Be$^+$-$^{40}$Ca$^+$) and two 
(for $^9$Be$^+$-$^{40}$Ca$^+$-$^9$Be$^+$) free parameters, see the main text.
$\omega_0 /(2\pi)=1.2$ MHz, $\gamma^2=3$.}
\end{center}
\end{figure}
% % % % % % % % % % % % % % % % % % % % % % % % % % % % % % % % % % % % % % % % % % %
% % % % % % % % % % % % % % % % % % % % % % % % % % % % % % % % % % % % % % % % % % %
%

For two different ions, we use a 13-th order polynomial $\rho_-(t)=\sum_{n=0}^{13}a_nt^n/t_f^n$, which is enough to nearly satisfy $\alpha_\pm(t_f)=\dot{\alpha}_\pm(t_f)=0$ and $\rho_+(t_f)=\gamma$ by finding suitable values for the four free parameters $a_{10-13}$.
As before, $\rho_+(t_f)=\gamma$ is nearly satisfied without any special design. Fig. \ref{excdiff} shows the final excitation for
a chain of two different ions. The excitation is higher than for equal masses. 
Both for the equal mass and different mass expansions, the (exact) excitation energy increases at short times, where the quadratic approximation to set the NM Hamiltonians fails, see Figs. \ref{optexc} and \ref{excdiff}.
%At longer times, where the approximation is valid, the approximate energy in Eq. (\ref{energy}) coincides with the energy in the laboratory frame. In this limit, the minimization we are doing in the NM frame straightfordwardly works in the full Hamiltonian.
Further simulations indicate that the larger the ratio between the masses, the higher the excitation.

%
%
% % % % % % % % % % % % % % % % % % % % % % % % % % % % % % % % % % % % % % % % %
% % % % % % % % % % % % % % % % % % % % % % % % % % % % % % % % % % % % % % % %
\begin{figure}[t]
\begin{center}
\includegraphics[width=8cm]{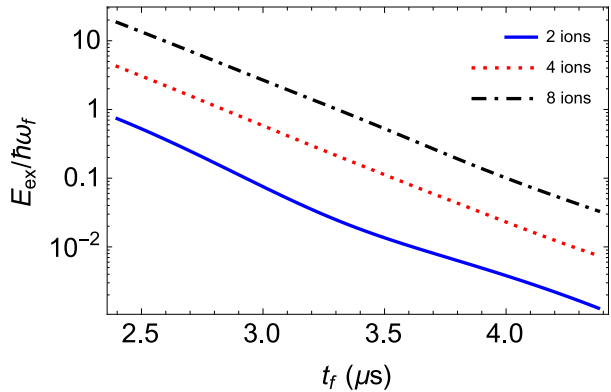}
\caption{\label{exc}(Color online)
Final excitation vs. final times for expansions of two equal ions (solid blue line), 4 equal ions (red dots) and
8 equal ions (dash-dotted black line). The simulations are performed according to the approximate protocol in Eq. (\ref{omega}) 
and by solving the classical equations of motion for $^{40}$Ca$^+$ ions. The initial ion chain is at equilibrium.   
$\omega_0 /(2\pi)=1.2$ MHz, $\gamma^2=3$.}
\end{center}
\end{figure}
% % % % % % % % % % % % % % % % % % % % % % % % % % % % % % % % % % % % % % % % % % %
% % % % % % % % % % % % % % % % % % % % % % % % % % % % % % % % % % % % % % % % % % %
%

A less accurate, approximate treatment is based on the
simpler polynomial ansatz $\rho_- =\sum_{n=0}^9a_nt^n$ without free parameters,\footnote{As in the transport of two ions \cite{Palmero},
an alternative ansatz to the polynomial is
$
\rho_-(t)=\frac{1+\gamma}{2}+\frac{\gamma -1}{256}\sum_{n=1}^3a_n\cos\left(\frac{(2n-1)\pi t}{tf}\right),
$
where $a_n=(-150,25,-3)$. In numerical calculations the polynomial ansatz (\ref{protocol}) performs slightly better than the cosine-based one.}
\beqa
\label{protocol}
\rho_- &=&126(\gamma -1)s^5-420(\gamma -1)s^6+540(\gamma -1)s^7
\nonumber\\
&-&315(\gamma -1)s^8+70(\gamma -1)s^9+1,
\eeqa
$s=t/t_f$. While the BC of $\rho_+$ and $\alpha_\pm$ are in general not accounted for exactly,
an advantage of this procedure is that there is no need to perform any numerical minimization. This is useful to generalize the method for larger ion chains.
For equal masses, both $\alpha_-=0$ and $\rho_-(t)$ are correctly designed, so that the center of mass
is not excited.
From Eq. (\ref{auxiliaryexp1}),  $\omega_1(t)$ is given by
%needs to follow
%in order to suppress excitations in both modes after expanding the trap from $\omega_0$ to $\omega_f$
%
\beq
\label{omega}
\omega_1=\sqrt{\frac{\omega_0^2}{\rho_-^4}-\frac{\ddot{\rho}_-}{A_-^2\rho_-}},
\eeq
where $A_-=\Omega_-/\omega_1$ is a constant, see Eq. (\ref{ome}).
In Fig. \ref{optexc} we  compare the performance of this approximate protocol
and the  one that satisfies all the BC in the two-equal-ion expansion. 
%Fig. 3 shows the values of the free parameters that minimize the excitation in the expansion of equal ions.
%
%
% % % % % % % % % % % % % % % % % % % % % % % % % % % % % % % % % % % % % % % % %
% % % % % % % % % % % % % % % % % % % % % % % % % % % % % % % % % % % % % % % %
\begin{figure}[t]
\begin{center}
\includegraphics[width=8cm]{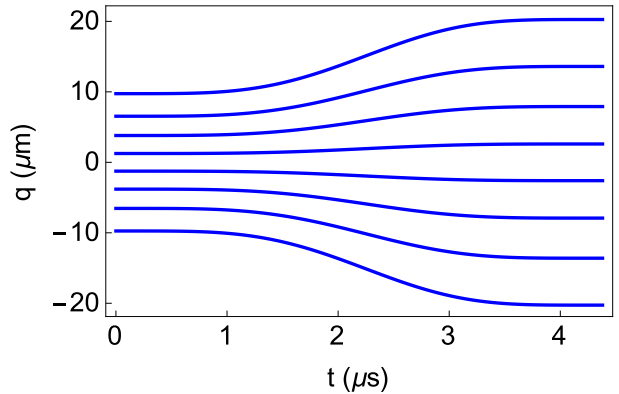}
\caption{\label{trajectories}(Color online)
Classical trajectories of eight expanding $^{40}$Ca$^+$ ions. The evolution is performed according to Eq. (\ref{omega}).
$\omega_0 /(2\pi)=1.2$ MHz, $\gamma^2=3$, $t_f=4.4$ $\mu$s.}
\end{center}
\end{figure}
% % % % % % % % % % % % % % % % % % % % % % % % % % % % % % % % % % % % % % % % % % %
% % % % % % % % % % % % % % % % % % % % % % % % % % % % % % % % % % % % % % % % % % %
%
%
%
%
%
%
\section{N-ion chain expansion\label{Nion}}
We now proceed to extend the results in the previous section to larger ion chains
governed by the  Hamiltonian (\ref{exphamiltonian}).
The equilibrium positions can be written  in the form \cite{James}
\beq
q_i^{(0)}(t)=l(t)u_i,
\eeq
where
\beq
l^3(t)=\frac{C_c}{u_0(t)}
\eeq
and the $u_i$ are the solutions of the system
\beq
u_i-\sum_{j=1}^{i-1}\frac{1}{(u_i-u_j)^2}+\sum_{j=i+1}^N\frac{1}{(u_i-u_j)^2}=0.
\eeq
The NM coordinates are thus defined as \cite{13Home}
\beq
{\sf q}_\nu =\sum_ia_{\nu i}\sqrt{m_j}(q_i-q_i^{(0)}),
\eeq
where the NM subscript $\nu$ runs now from $1$ to $N$.  Conventionally the
$\nu$ are ordered from the lowest to the highest frequency \cite{James}.  
As for two ions we define $V(q_1,q_2,q_3,...,q_N)$ as
the coordinate-dependent part of  the Hamiltonian (\ref{exphamiltonian}). The $a_{\nu i}$ are the components of the 
$\nu$-th eigenvector of the symmetric matrix 
$V_{ij}~=~\frac{1}{\sqrt{m_im_j}}\frac{\partial^2V}{\partial q_i\partial q_j}(q_i^{(0)},q_j^{(0)})$,
that, together with the eigenvalues $\lambda_\nu=\Omega_\nu^2$  will usually be determined numerically \cite{James}.
They are normalized as $\sum_i a_{\nu i}^2=1$. As $u_0$ is common to all ions, it can be shown that $\Omega_\nu(t)=A_\nu \omega_1(t)$, 
where $A_\nu$ is a constant. 
 
%
%
% % % % % % % % % % % % % % % % % % % % % % % % % % % % % % % % % % % % % % % % %
% % % % % % % % % % % % % % % % % % % % % % % % % % % % % % % % % % % % % % % %
\begin{figure*}[t!]
\begin{center}
\includegraphics[width=8cm]{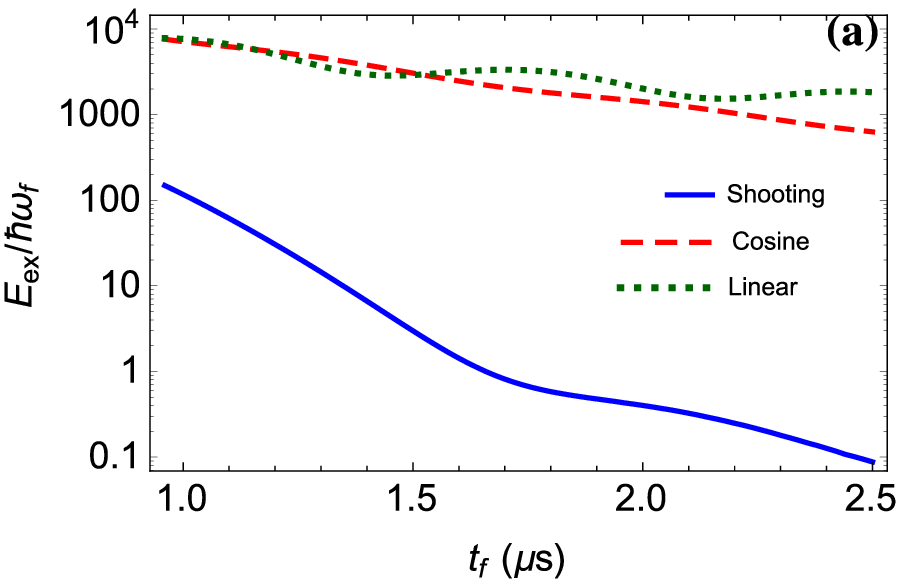}
\includegraphics[width=8cm]{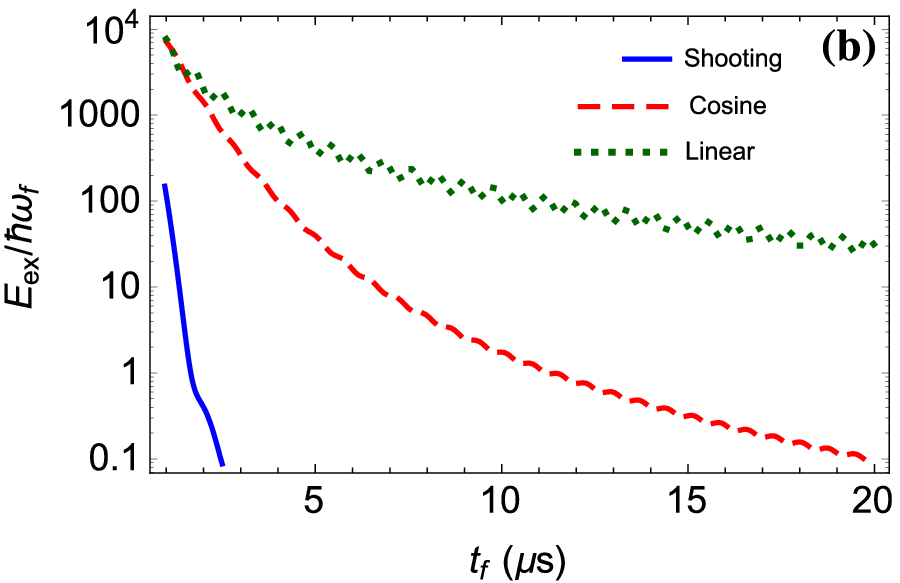}
\caption{\label{comparison}(Color online)
Comparison of final excitation quanta (classical simulation as in Fig. 1) 
vs final time in the expansion of 2 $^{40}$Ca$^+$ ions following the shooting protocol (solid blue), linear protocol in Eq. (\ref{linear}) (dotted green) and cosine protocol  in Eq. (\ref{cosine}) (dashed red). In a) we plot in logarithmic scale up to times where only the shooting protocol reaches the level of $0.1$ excitation quanta. 
In b) we extend the analysis up to longer final times, so that the best of the cosine protocol  reaches also 0.1 final excitation quanta.
$\omega_0 /(2\pi)=1.2$ MHz, $\gamma^2=3$.}
\end{center}
\end{figure*}
% % % % % % % % % % % % % % % % % % % % % % % % % % % % % % % % % % % % % % % % % % %
% % % % % % % % % % % % % % % % % % % % % % % % % % % % % % % % % % % % % % % % % % %
%%
%
Generalizing the steps leading to Eq. (\ref{Hprime}), the Hamiltonian in a NM frame up to quadratic terms becomes
\beq
H'=\sum_\nu \bigg[\frac{{\sf p}_\nu^2}{2}+\frac{1}{2}\Omega_\nu^2 {\sf q}_\nu^2 +{\sf p}_{0\nu}{\sf p}_\nu\bigg],
\eeq
where the ${\sf p}_\nu$ are momenta conjugate to the ${\sf q}_\nu$, and
${\sf p}_{0\nu}=-\sum_i a_{\nu i}\sqrt{m_i}\dot{q}_i^{(0)}$. As for two ions, all the ${\sf p}_{0\nu}$ are proportional to
$\dot{\omega}_1/\omega_1^{5/3}$.
We now apply the unitary transformation $\mathcal{U}=e^{-i\sum_\nu {\sf p}_{0\nu}{\sf q}_\nu /\hbar}$ and find 
the effective Hamiltonian
\beq
H''=\sum_\nu \bigg[\frac{{\sf p}_\nu^2}{2}+\frac{1}{2}\Omega_\nu^2\left({\sf q}_\nu+\frac{\dot{{\sf p}}_{0\nu}}{\Omega_+^2}\right)^2\bigg].
\eeq
This Hamiltonian is similar to the one for two ions  (\ref{hamNM}). The corresponding set of auxiliary equations
is also similar to Eqs. (\ref{auxiliaryexp1}) and (\ref{auxiliaryexp2}),
\beqa
\ddot{\rho}_\nu+\Omega_\nu^2\rho_\nu=\frac{\Omega_{0\nu}^2}{\rho_\nu^3},
\nonumber\\
\ddot{\alpha}_\nu+\Omega_\nu^2\alpha=\dot{{\sf p}}_{0\nu}.
\eeqa
The BC for inverse engineering read $\rho_\nu(0)=1$, $\rho_\nu(t_f)=\gamma$, $\dot{\rho}_\nu(t_b)=\ddot{\rho}_\nu(t_b)=0$, $\alpha_\nu(t_b)=\dot{\alpha}_\nu(t_b)=\ddot{\alpha}_\nu(t_b)=0$. When introducing the BC for the $\alpha_\nu$ in the set of Newton's equations, we get from all of them the same condition $\frac{\dot{\omega}_1(t_b)}{\omega_1(t_b)}+\ddot{\omega}_1(t_b)=0$, which is satisfied for $\dot{\omega}_1(t_b)=\ddot{\omega}_1(t_b)=0$.
%Also, when trying to solve for the Ermakov equations, we are only able to solve for one of the modes at the same time. Thus, the same protocols used for 2-ion expansion (\ref{omega}) should scale for any arbitrary number of ions.

Fig. \ref{exc} depicts the excitation for expansions of  single-species ion-chains, with  approximate (non-optimized) protocols 
that use Eqs. (\ref{protocol}) and (\ref{omega}), but with the lowest frequency mode, $\nu=1$, instead of the 
$minus$ $(-)$ mode.   
%The simulations were performed by following a classical evolution
%rather than the quantum evolution, which would be computationally very demanding.
%For the classical simulations, we solve Hamilton's equations rather than the Schr\"odinger equation, and start with
%the system at the potential minimum (zero energy). This  reproduces accurately the excitation energy of the quantum system
%as the potential is effectively nearly harmonic. Moreover, the evolution of the wave packet's width ($\rho_\pm$) is close to being adiabatic.
%so squeezing effects are irrelevant.
%
%Fig. \ref{optexc} shows the good agreement between quantum and classical simulations for  the two-ion expansion.
%For longer chains we only perform classical computations.
The longer the chain the lower the fidelity of the protocol, as more terms are neglected in the NM approximation and more boundary conditions are disregarded. However, the protocol still provides little
excitation %for short times which are relevant for quantum computation 
at long enough final times
in the most demanding simulation that we examined, $N=8$.
Fig. \ref{trajectories} shows the position of the ions, and the trap frequency along the evolution time for the eight-ion chain, ending up with a separation between ions twice as large as the initial one, in times shorter than 4 $\mu$s (Fig. \ref{exc}) without any significant final excitation.

In Fig. \ref{excdiff}  the excitation for an expansion of the two-species chain $^9$Be$^+$-$^{40}$Ca$^+$-$^9$Be$^+$ is depicted.
The minimization technique was used with two free parameters, that is, with an 11-th order polynomial ansatz for  $\rho_{\nu=1}(t)$.   
The excitation is smaller than for the shorter chain $^9$Be$^+$-$^{40}$Ca$^+$ (with a 13-th order polynomial for $\rho_-$) 
due to the symmetry in the three-ion chain, which leaves two of the NM static and unexcited.

\section{Discussion}
%
%
%
%
%
%Several aspects of non-adiabatic excitation-free expansions have been investigated,
%such as optimal design with respect to
%different criteria \cite{Ste1,Ste2,Ste3}, transient excitation \cite{energy},
%or three-dimensional effects \cite{3D}.
%
We have designed fast diabatic protocols for the time dependence of the trap frequency that suppress the final excitation of different ion-chain expansions or compressions. Unlike the simpler
single-ion expansion \cite{Xi}, the inverse design problem of the trap frequency for an
ion chain involves 
coupled Newton and Ermakov equations for each dynamical normal mode.
We found ways to deal with this
inverse problem by applying
a shooting technique in the most accurate protocols, and effective, simplifying approximations.
%The analysis is applicable to for arbitrarily long ion chains, and for chains containing any sequence of either single or mixed-species ions.

%
These protocols work for process times for which the quadratic approximation
for the Hamiltonian
is valid. Longer and more asymmetric chains need larger times than shorter and symmetrical ones.
The examples show that these times are compatible with current quantum information protocols,
%of relevance for quantum information processing and
%short enough to avoid decoherence, 
so many processes may benefit
by the described trap frequency time dependencies.

The designed protocols provide a considerable improvement in final time and excitation energy with respect to simple, naive protocols. 
For the expansion of two $^{40}$Ca$^+$ considered in Fig. 1 we compare in Fig. \ref{comparison} the excitation energy of the shooting protocol with two simple protocols that drive the frequency $\omega_1$  linearly,
\beq
\label{linear}
\omega_1(t)=\omega_0+\frac{\omega_f-\omega_0}{t_f}t,
\eeq
and following a cosine function, 
% The linear protocol is given by
\beq
\label{cosine}
\omega_1(t)=\frac{\omega_0+\omega_f}{2}+\frac{\omega_0-\omega_f}{2}\cos\left(\frac{\pi t}{t_f}\right).
\eeq
%
%The cosine protocol, additionally satisfies that its first derivative is 0 both at initial and final times. One can thus expect a better performance than for the linear protocol, Fig. \ref{comparison}. Nevertheless, 
The simulations are classical, as described in Sec. \ref{2ion}.   
%The shooting method clearly outperforms the simple ones, as shown in  Figs. \ref{comparison} (a) and (b).   
Fig. \ref{comparison} (a) compares the excitations at short times. For  $t_f\sim 2.5$ $\mu$s,  the shooting protocol reaches a 
low excitation of 0.1 vibrational quanta, four orders of magnitude smaller than the excitations due to the simple methods. 
In Fig. \ref{comparison} (b) the excitations are represented for longer protocol times. The smoother cosine protocol 
behaves better than the linear one and finally reaches an excitation of approximately 0.1 quanta for $t_f\sim 20$ $\mu$s. 
%That means that the optimised protocol gets an improvement of 8 times shorter final time.

%
%
%
%
\section*{Acknowledgements}
We thank Ryan Bowler, John Gaebler, and Dietrich Leibfried for discussions. 
This work was supported by the Basque Country Government (Grant No. IT472-10), Ministerio de Econom\'\i a y Competitividad (Grant No. FIS2012-36673-C03-01), the program UFI 11/55 of UPV/EHU, the SNF under grant number COST-C12.0118, and the ETH Zurich.
M.P. and S.M.-G. acknowledge  fellowships by UPV/EHU. M.P. is also grateful to the STSM fellowship program by the COST action IOTA.

\end{document}